\documentclass[aps,prd,nofootinbib,showpacs,tightenlines,preprint,eqsecnum,amsmath,amsfonts]{revtex4}
\usepackage{bm}

\newcommand{\be}{\begin{equation}}
\newcommand{\ee}{\end{equation}}
\newcommand{\bea}{\begin{eqnarray}}
\newcommand{\eea}{\end{eqnarray}}

\def\beq{\begin{equation}}
\def\eeq{\end{equation}}
\def\ds{\displaystyle}

\def\req#1{(\ref{#1})}
\def\text#1{\mbox{\scriptsize{#1}}}
\def\B{\mbox{B}}

\begin{document}

\title{Casimir Energy for a Purely Dielectric Cylinder by the Mode
Summation Method}
\author{August Romeo}
\affiliation{}

\author{Kimball A. Milton}
\email{milton@nhn.ou.edu}
\affiliation{Oklahoma Center for High Energy Physics and Department of
Physics and Astronomy, University of Oklahoma, Norman, OK 73019, USA}

\date{\today}

\begin{abstract}
We use the mode summation method together with zeta-function regularization
to compute the Casimir energy of a dilute dielectric cylinder.  The
method is very transparent, and sheds light on the reason the resulting
energy vanishes.
\end{abstract}


\maketitle
\section{Motivation}
A few years ago a calculation of the sum of
van der Waals interactions for a purely dielectric cylinder
in the
dilute-dielectric approximation led to a surprising null result \cite{MNN}.
(This was first found by an unpublished calculation by the first author
of the present paper, and later independently confirmed by calculations
by Milonni \cite{milonni} and by Barton \cite{barton}.)
This unexpected finding produced
a quest to calculate  the corresponding Casimir energy with the aim of
verifying the predicted equality between both quantities, which was
recently established in Ref.~\cite{CPM}.  However, a physical understanding
of why this Casimir energy should vanish (as also does that for a
dilute dielectric-diamagnetic cylinder \cite{MNN}) remains elusive.
In addition to the physical interest of this subject, we recall that
 the authors of Ref.~\cite{MNN} commented at some length upon the
comparative advantages and shortcomings of Green's function formalisms and
mode summation methods for the evaluation of Casimir energies.
A formal relation between these two approaches was
given in appendix A of that paper.

The first procedure is essentially a calculation of the expectation value
of the stress-energy tensor expressed in terms  of Green's functions and their
transforms. This approach has proven to be remarkably fruitful and enlightening
from the perspective of physical interpretation, as the
quantities computed bear close relationships to easily identifiable observables
expressed in terms of sources and fields.
This was the method applied in Ref.~\cite{CPM} to the object of our
present study, i.e., the Casimir energy for a dielectric cylinder with
the light velocity different on the inside and the outside.

In contrast, the mode summation method rests on the concept of zero point
energy and its basic expression as an infinite sum of
eigenfrequencies. Even though this sum is
a rather abstruse concept, from a prosaic viewpoint it has
the appeal of simplicity. Since these eigenfrequencies stem
from classical problems and are in many cases already known
(e.g. through textbooks like Ref.~\cite{S}), the only remaining task is
to perform their summation. Once this question has been
mathematically posed, the  answer sought after is rather
easily within one's grasp.

In view of these considerations, the reflections in
Ref.~\cite{MNN} itself, and the examples offered by many related
works, it is sensible to say that the two procedures may
complement each other.  This is especially so, since these calculations
still present a number of subtleties, and it is to be hoped that approaching
the problems from disparate viewpoints may lead to improved physical
insight.  Specifically, the well-known divergence difficulties,
which have recently received much attention \cite{jaffe}, encourage us
to examine the problem anew.
With this idea in mind, we shall tackle some aspects of the problem treated
in Ref.~\cite{CPM} by a particular variant of the mode summation method.

\section{Mode sum}

According to Refs.~\cite{S,MNN}, the eigenfrequencies $\omega_{m,p,k_z}$ of the
Maxwell equations
for an infinite material cylinder of radius $a$, oriented along the $z$-axis,
with permittivity and permeability $(\varepsilon_1, \mu_1)$,
surrounded by a medium
with permittivity and permeability $(\varepsilon_2, \mu_2)$,
are the zeros of the following function:
\begin{subequations}
\beq
f_m(k_z,\omega)=0,\quad m=0,\pm1,\pm2,\dots,\quad k_z\in\mathbb{R},
\label{fn1}
\eeq
\beq
f_m(k_z,\omega)
\equiv \frac{1}{\Delta^2}\left[
\Delta_m^{\text{TE}}(x, y)\, \Delta_m^{\text{TM}}(x, y)
-m^2\frac{a^4 \omega^2 k^2_z }{ x^2 y^2 }
(\varepsilon_1 \mu_1-\varepsilon_2 \mu_2 )^2
J_m^2(x)\,H_m^2(y)
\right],
\label{fn2}
\eeq
\end{subequations}
with the following abbreviations
\beq
\begin{array}{c}
\ds \Delta= -\frac{2i }{ \pi}, \quad x= \lambda_1 a, \quad y= \lambda_2 a, \\
\begin{array}{rcl}
\Delta_m^{\text{TE}}(x, y)&=&\ds
\mu_1 y J_m'(x)\,H_m(y)-
\mu_2 x J_m(x)\,H'_m(y), \\
\Delta_m^{\text{TM}}(x, y)&=&\ds
\varepsilon_1 y J_m'(x)\,H_m(y)-
\varepsilon_2 x J_m(x)\,H'_m(y),
\end{array} \\
\lambda_i^2=\varepsilon_i \mu_i \omega^2-k_z, \quad i=1,2.
\end{array}
\label{defsDeltas}
\eeq
As usual, $m$ is the azimuthal quantum number,
 $k_z$ denotes the momentum along the axis of the cylinder, and
 $p$ labels the zeroes of $f_m(k_z,\omega)$. Here, for
$y>0$, $H_m(y)\equiv H_m^{(1)}(y)$.
Note that $f_m= -\Delta^{-2}\Xi$, where $\Xi$ is the same
denominator introduced in Ref.~\cite{CPM} and the $\Delta^{-2}$ factor has been
introduced for convenience.
Recalling that the velocities of light in each media are given by
$c_i= (\varepsilon_i\mu_i)^{-1/2}$, $i=1,2$, we see that
Eqs.~\req{fn1}, \req{fn2} exhibit a peculiar feature of this situation:
if $c_1 \neq c_2$, the second term in \req{fn2}
comes into play, and it is no longer possible to decompose the mode set
into zeros of $\Delta_m^{\text{TE}}$ and zeros of $\Delta_m^{\text{TM}}$; that
is, the transverse electric and magnetic modes become entangled.
When medium 1 is purely dielectric and medium 2 is vacuum,
$\varepsilon_1= \varepsilon$, $\mu_1=1$,
$\varepsilon_2=\mu_2= 1$, one may write
$\omega=a^{-1}(y^2+\widehat k^2)^{1/2}$ and
$x^2=y^2+(\varepsilon-1)(y^2+\widehat k^2)$,
where
$\widehat{k}\equiv k_z a$.  We will also set $c_2=c=1$.

Formally speaking, the Casimir energy per unit length is given
by the mode sum
\beq
\mathcal{E}_C= \frac{1 }{ 2}\hbar
\int_{-\infty}^{\infty} \frac{d k_z}{ 2\pi}
\sum_{m, p} \omega_{m,p,k_z}.
\label{eC}
\eeq
This quantity is related to the Casimir radial pressure $P_C$ through
\beq
P_C=\frac{1 }{ \pi a^2} \mathcal{E}_C.
\label{PepsC}
\eeq
The sum in \req{eC} is divergent and will be regularized by changing the
power of $\omega$, which power becomes the argument of the associated
zeta function. By Ref.~\cite{BP} we already
know that, up through the order of $(\varepsilon-1)^2$,
there should be no ambiguities, i.e., no logarithms of
arbitrary scales, because the heat kernel coefficient which would
multiply them is of ${\cal O}((\varepsilon-1)^3)$. Therefore, we are permitted
to define
\beq
\mathcal{E}_{C}(s)= \frac{\hbar}{ 2} \,
\int_{-\infty}^{\infty} \frac{d k_z}{ 2\pi} \,
\sum_{m, p}
\omega_{m, p, k_z}^{-s}
=\frac{\hbar}{ 2} \,  a^{s-1}
\int_{-\infty}^{\infty} \frac{d \widehat k }{ 2\pi} \,
\sum_{m, p} ( y_{m, p}^2+\widehat k^2 )^{-\frac{s }{ 2}},
\label{zetaOm}
\eeq
as a function of the complex variable $s$, without any additional mass
scale.
It is proposed to make sense of Eq.~\req{eC} by means of analytic
continuation of this function to $s= -1$, namely,
\beq
\mathcal{E}_C= \lim_{s\to -1} \mathcal{E}_C(s).
\label{eCz}
\eeq
For any given values of  $\widehat k$, $m$, the remaining sum over $p$
may be rewritten, with the help of the residue theorem,
as a contour integral in complex $y$ plane. Thus,
\beq
\mathcal{E}_{C}(s)= \frac{\hbar}{ 2} \,
a^{s-1}
\int_{-\infty}^{\infty} \frac{d \widehat k }{ 2\pi}
\sum_{m=-\infty}^{\infty} \,
\frac{s }{ 2\pi i} \,
\int_{C} \, dy \, y \, (y^2+\widehat k^2)^{-\frac{s+2 }{ 2}} \, \ln f_m,
\label{summ}
\eeq
where $C$ is a circuit which encloses all the $y$ values corresponding
to the positive zeroes of $f_m$.  This approach is often referred to as
the argument principle \cite{schram}.
Quite often in the application of this method, an asymptotic form
$f_{m, \text{as}}$ of $f_m$ is found and then $\ln f_{m, \text{as}}$ is
subtracted from $\ln f_m$ in the integrand.
Actually, the factors introduced in \req{fn2} relative
to the original $f_m$ of Ref.~\cite{MNN} amount to dividing
that function by the leading term of its asymptotic behaviour.
At any rate, the full form of this asymptotic behaviour is presumably
related to the limit of unbounded space, which is already available as
the bulk contribution calculated in Ref.~\cite{CPM}.

Next, the dilute-dielectric approximation will be made by
expanding the logarithm function of Eq.~\req{summ} in powers of
$(\varepsilon-1)$, choosing $y$ as the independent variable and
taking $x$ as a function of $y$ and $\widehat k$:
\beq
\begin{array}{ll}
\ln f_m
=&\ds\left[
L_{m 1}^{0}(y)
+L_{m 1}^{1}(y)(y^2+\widehat{k}^2)
\right] (\varepsilon-1) \\
&\ds+ \left[
L_{m 2}^{00}(y)
+L_{m 2}^{10}(y)(y^2+\widehat{k}^2)
+L_{m 2}^{20}(y)(y^2+\widehat{k}^2)^2
+ L_{m 2}^{11}(y)(y^2+\widehat{k}^2) \, \widehat{k}^2
\right](\varepsilon-1)^2 \\
&\ds\mbox{}+{\cal O}((\varepsilon-1)^3),
\end{array}
\label{expln}
\eeq
where
\beq
\begin{array}{lll}
L_{m 1}^{0}(y)&=&\ds \frac{1 }{ \Delta} y \, J_m'(y) H_m(y), \\
L_{m 1}^{1}(y)&=&\ds \frac{1 }{ \Delta y} \Delta_m^{(1,0)}(y) , \\
L_{m 2}^{00}(y)&=&\ds -\frac{1 }{ 2\Delta^2} y^2 \, {J_m'}^2(y) H_m^2(y), \\
L_{m 2}^{10}(y)&=&\ds -\frac{1 }{ 2\Delta^2} \left[
\Delta_m^{(1,0)}(y) J_m'(y) H_m(y)
+ \frac{\Delta}{ y} \left(
J_m'(y)+y \left( 1 -\frac{m^2}{ y^2} \right) J_m(y)
\right) H_m(y)
\right], \\
L_{m 2}^{20}(y)&=&L_{m 2}^{20 A}(y)+L_{m 2}^{20 B}(y) , \quad
\left\{
\begin{array}{lll}
L_{m 2}^{20 A}(y)&=&\ds \frac{1 }{ 4 \Delta y^2}
\left( \Delta_m^{(2,0)}(y)-\frac{1 }{ y}\Delta_m^{(1,0)}(y) \right) , \\
L_{m 2}^{20 B}(y)&=&\ds -\frac{1 }{4 \Delta^2 y^2}
\left( \Delta_m^{(1,0)}(y) \right)^2 ,
\end{array}
\right. \\
L_{m 2}^{11}(y)&=&\ds-\frac{m^2 }{ \Delta^2 y^4} J_m^2(y) H_m^2(y).
\end{array}
\label{Ls}
\eeq
Here, the notation is
\beq
\Delta_m^{(j,0)}(y)=
\left.
\frac{\partial^j \Delta_m(x,y)}{ \partial x^j}
\right\vert_{\varepsilon_1=\varepsilon_2=1, \mu_1=\mu_2=1, x=y},
\mbox{ for $j= 1,2$},
\eeq
where $\Delta_m$ stands for
either $\Delta_m^{\text{TE}}$ or $\Delta_m^{\text{TM}}$
(in the free space limit there is no difference). Moreover, note that
\beq
\Delta_m(x,y)|_{\varepsilon_1=\varepsilon_2=1, \mu_1=\mu_2=1, x=y}
=-y \, W[J_m(y), H_m(y)]
=\Delta ,
\eeq
where $W$ denotes the Wronskian.
Explicitly,
\beq
\begin{array}{lll}
\Delta_m^{(1,0)}(y)&=&\ds
-\frac{1 }{ y}\left[ y^2 J_m'(y) H_m'(y)+(y^2-m^2) J_m(y)H_m(y) \right]
-(J_m(y)H_m(y))' , \\
\Delta_m^{(2,0)}(y)&=&\ds \left( \Delta_m^{(1,0)}(y) \right)'
-\left(1-\frac{m^2+1 }{ y^2}\right)\Delta,
 \end{array}
\label{Deltans}
\eeq
where
$\ds(\Delta_m^{(1,0)}(y))' \equiv \frac{d }{ dy}\Delta_m^{(1,0)}(y) \neq
\Delta_m^{(2,0)}(y)$, as shown.
Once this task has been accomplished,
we insert Eq.~\req{expln} into Eq.~\req{summ} and perform the $\widehat k$
integration.

 Consideration of formulas
\req{summ}--\req{Deltans}
shows the need of dealing with integrals of the type
\beq
I\equiv \int_{-\infty}^{\infty} d\widehat k
\int_C dy \, y \, F(y) \, (y^2+\widehat k^2)^{-\alpha} \,
\widehat k^{2\beta} ,
\label{CI0}
\eeq
where $C$ is the contour specified above, and $F$
satisfies $F(-iv)= F(iv)$ for $v\in{\mathbb R}$,
as well as having adequate asymptotic properties (see below).
Looking at the $(y^2+\widehat k^2)$ powers in Eqs.~\req{summ}, \req{expln},
we see that in the required
cases $\alpha= s/2+1, s/2, s/2-1$, and $\beta=0$
except for one integral with $\beta=1$.
Since $\alpha$ is just a translation of $s/2$, analytic continuation
in $s$ amounts to analytic continuation in $\alpha$.

Straightforward integration for $\widehat k$ yields
\beq
I= \B\left( \beta+\frac{1 }{ 2}, \alpha-\beta-\frac{1 }{ 2} \right)
\int_C dy \, y^{2-2\alpha+2\beta} \, F(y) ,
\label{CI2}
\eeq
where $\B$ stands for the Euler beta function
$\ds\B(x,y)= \frac{\Gamma(x) \Gamma(y) }{\Gamma(x+y)}$.
Now we rotate the $C$  contour so that it
consists of a straight line parallel to and just to the right
of the imaginary axis, closed by a semicircle of infinitely large
radius on the right.\footnote{Another way to describe this rotation
is in terms of a purely mathematical transformation, based on the
required analyticity of the underlying Green's function.  See
Refs. \cite{Milton:1978sf,Brevik:2000hk}.}
 The branch line of the $y^{2-2\alpha+2\beta}$
function, which starts at the origin, is placed so
that the circuit does not cross it, that is, it lies
along the negative real axis.
(In the limit where the vertical part of $C$
overlaps the axis, the origin could be avoided by a small semicircle, and,
eventually, the integration along this infinitesimal part would vanish as
 $s\to-1$).
Further, it is assumed that, combining
the asymptotic behaviour of $F$ and the possibility of varying
the $\alpha$ value as necessary, the contribution from the large
semicircle vanishes when its radius tends to infinity.
Thus, the $y$ integral reduces to an integration along the vertical
parts of $C$, where $y=\pm iv$, and one has
$y^{2-2\alpha+2\beta}=e^{\pm i\pi(1-\alpha+\beta)} v^{2-2\alpha+2\beta}$
on the upper and lower segments.
As a result,
\beq
I= -2i \, \B\left( \beta+\frac{1 }{ 2}, \alpha-\beta-\frac{1 }{ 2} \right)
\sin\left( \pi \left( \alpha-\beta-\frac{1 }{ 2} \right) \right)
\int_{0}^{\infty} dv \, v^{2-2\alpha+2\beta} \, F(iv) .
\label{CIpre2}
\eeq
Applying the reflection formula
$
\ds\Gamma(z) \Gamma(1-z)=\frac{\pi }{\sin\pi z}
$
to the gamma functions in the Euler beta function, we re-express $I$  as
\beq
I= -2i \, \B\left( \beta+\frac{1 }{ 2}, 1-\alpha \right) \sin(\pi\alpha)
\int_{0}^{\infty} dv \, v^{2-2\alpha+2\beta} \, F(iv) .
\label{CI1}
\eeq
Note that for the $\alpha$ values corresponding to $s=-1$
($\alpha= 1/2, -1/2, -3/2$), and for $\beta=0,1$,
the pole of the $\B$ function and the zero of the sine function
in Eq.~\req{CIpre2} combine to give a finite product,
while in Eq.~\req{CI1} each of these two factors
is already finite.

Applying formula \req{CI1} to Eqs.~(\ref{summ}), (\ref{expln})
yields the following result:
\beq
\mathcal{E}_C(s)
= \mathcal{E}_{C 1} (\varepsilon-1) + \mathcal{E}_{C 2}(s) (\varepsilon-1)^2
+{\cal O}((\varepsilon-1)^3),
\label{epsc12exp}
\eeq
where
\beq
\begin{array}{c}
\mathcal{E}_{C 1}(s)=
\mathcal{E}_{C 1}^{0}(s)
+\mathcal{E}_{C 1}^{1}(s), \\
\left\{
\begin{array}{lll}
\ds \mathcal{E}_{C 1}^{0}(s)&=&\ds-\frac{\hbar}{ 2} \, \frac{s a^{s-1} }{ 2\pi^2} \
\B\left( \frac{1 }{ 2}, -\frac{s }{ 2} \right)
\sin\left( -\pi \frac{s}{ 2} \right)
\sum_{m=-\infty}^{\infty}\int_{0}^{\infty} dv \, v^{-s} L_{m 1}^{0}(iv) , \\
\ds \mathcal{E}_{C 1}^{1}(s)&=&\ds -\frac{\hbar}{ 2} \, \frac{s a^{s-1} }{ 2\pi^2}
\ \B\left( \frac{1 }{ 2}, 1-\frac{s}{ 2} \right)
\sin\left( \pi\frac{s}{ 2}  \right)
\sum_{m=-\infty}^{\infty}\int_{0}^{\infty} dv \, v^{2-s} L_{m 1}^{1}(iv) ,
\end{array}
\right.
\end{array}
\label{epsc1}
\eeq
and
\beq
\begin{array}{c}
\mathcal{E}_{C 2}(s)=
\mathcal{E}_{C 2}^{00}(s)
+\mathcal{E}_{C 2}^{10}(s)
+\mathcal{E}_{C 2}^{20 A}(s)+\mathcal{E}_{C 2}^{20 B}(s)
+\mathcal{E}_{C 2}^{11}(s), \\
\left\{
\begin{array}{lll}
\ds \mathcal{E}_{C 2}^{00}(s)&=&\ds -\frac{\hbar}{ 2} \, \frac{s a^{s-1} }{ 2\pi^2} \
\B\left( \frac{1 }{ 2}, -\frac{s}{ 2} \right)
\sin\left( -\pi \frac{s}{ 2} \right)
\sum_{m=-\infty}^{\infty}\int_{0}^{\infty} dv \, v^{-s} L_{m 2}^{00}(iv) , \\
\ds \mathcal{E}_{C 2}^{10}(s)&=&\ds -\frac{\hbar }{ 2} \, \frac{s a^{s-1} }{2\pi^2} \
\B\left( \frac{1 }{ 2}, 1-\frac{s}{ 2} \right)
\sin\left( \pi \frac{s}{ 2}  \right)
\sum_{m=-\infty}^{\infty}\int_{0}^{\infty} dv \, v^{2-s} L_{m 2}^{10}(iv) , \\
\ds \mathcal{E}_{C 2}^{20 A,B}(s)&=&\ds -\frac{\hbar }{ 2} \, \frac{s a^{s-1} }{ 2\pi^2} \
\B\left( \frac{1 }{ 2}, 2-\frac{s}{ 2} \right)
\sin\left(- \pi \frac{s}{ 2} \right)
\sum_{m=-\infty}^{\infty}\int_{0}^{\infty} dv \, v^{4-s} L_{m 2}^{20 A,B}(iv) , \\
\ds \mathcal{E}_{C 2}^{11}(s)&=&\ds -\frac{\hbar }{ 2} \, \frac{s a^{s-1}}{2\pi^2} \
\B\left( \frac{3 }{ 2}, 1-\frac{s}{ 2} \right)
\sin\left( \pi \frac{s}{ 2}  \right)
\sum_{m=-\infty}^{\infty}\int_{0}^{\infty} dv \, v^{4-s} L_{m 2}^{11}(iv) .
\end{array}
\right.
\end{array}
\label{epsc2}
\eeq

\subsection{First order in $(\varepsilon-1)$}

Taking $\mathcal{E}_{C 1}^{0}(s)$ from \req{epsc1},
and $L_{m 1}^{0}(iv)$ from \req{Ls} one is led to
\beq
\mathcal{E}_{C 1}^{0}(s)=-\frac{\hbar}{ 2} \, \frac{s a^{s-1} }{ 2\pi^2} \
\B\left( \frac{1 }{ 2}, -\frac{s }{ 2} \right)
\sin\left( -\pi \frac{s}{ 2} \right)
\sum_{m=-\infty}^{\infty}\int_{0}^{\infty} dv \, v^{1-s}  \\
I_m'(v) K_m(v).
\eeq
The beta and sine functions are finite at $s=-1$. As for the
integral, it will be rewritten by introducing the factor
$1= -v W[I_m(v),K_m(v)]= -v [I_m(v) K_m'(v) - I_m'(v) K_m(v)]$ for each $m$:
\beq
\begin{array}{c}
\ds\int_{0}^{\infty} dv \, v^{1-s}
\sum_{m=-\infty}^{\infty} I_m'(v) K_m(v)= \\
\ds -\int_{0}^{\infty} dv \, v^{2-s}
\sum_{m=-\infty}^{\infty} I_m(v) I_m'(v) K_m(v) K_m'(v)
+\int_{0}^{\infty} dv \, v^{2-s}
\sum_{m=-\infty}^{\infty} {I_m'}^2(v) K_m^2(v).
\label{2.21}
\end{array}
\eeq

We carry out the summation over $m$ by using
the addition theorem for the modified Bessel functions:
\beq
\begin{array}{rcl}
\ds\sum_{m=-\infty}^{\infty}  I_m(kr) K_m(k\rho) \, e^{i m \phi}
&=&K_0( k R(r,\rho,\phi) ) \\
R(r,\rho,\phi)&=&\ds \sqrt{ r^2+\rho^2-2r\rho \cos\phi } ,
\quad \rho > r.
\end{array}
\label{thrm}
\eeq
Expressions for products
of four Bessel functions are found by
multiplying differentiated versions of the Bessel function addition
theorem \req{thrm},  integrating over $\phi$,
and setting $k r=k \rho \equiv v$.
A  change of variable $u= \sin\frac{\phi}{2}$ is made.
Next, instead of proceeding with the evaluation of the resulting $u$
integral, the expression is multiplied by a $v$ power involving $s$,
and integrated over $v$, with the help of
\beq
\int_{0}^{\infty} dx \, x^{-\lambda} K_{\mu}(x) K_{\nu}(x)=
2^{-2-\lambda}
\frac{\Gamma\left( \frac{1-\lambda+\mu+\nu }{2} \right)
 \Gamma\left( \frac{1-\lambda-\mu+\nu }{2 }\right)
 \Gamma\left(\frac{ 1-\lambda+\mu-\nu }{ 2 }\right)
 \Gamma\left( \frac{1-\lambda-\mu-\nu }{ 2 }\right)
 }{ \Gamma(1-\lambda) }
\eeq
(from result (6.576.4) in Ref.~\cite{GR}).
Then the $u$ integration is carried out afterwards.
Thus, the emerging expressions depend on $s$ only through gamma
functions (see also Refs.~\cite{KR,CPM}).
In fact, the results in formul\ae\ (81) of Ref.~\cite{CPM} may be viewed as
intermediate steps, the final result being for example
\bea
\ds \int_{0}^\infty dv \, v^{2-s} \sum_{m=-\infty}^{\infty} {I_m'}^2(v)
K_m^2(v)&=&
\ds \int_{0}^\infty dv \, v^{2-s} \sum_{m=-\infty}^{\infty} {K_m'}^2(v)
I_m^2(v)\nonumber\\=
\ds \int_{0}^\infty dv \, v^{2-s} \sum_{m=-\infty}^{\infty} I_m(v) I_m'(v)
K_m(v) K_m'(v)
&=& \ds \frac{1 }{ 8\pi^{1/2}}
\ds \frac{\Gamma\left(\frac{ 5-s }{ 2 }\right)
\Gamma^2\left(\frac{ 3-s }{ 2 }\right)
\Gamma\left( \frac{1-s }{ 2 }\right)
}{ \Gamma(3-s)}
\frac{\Gamma\left(\frac{ s }{ 2 }\right) }{
\Gamma\left(\frac{ s+1 }{2} \right) }. \nonumber\\\label{ins}
\eea
Even if the left hand side of each such integral
 is not initially defined for $s=-1$, the
right hand side provides the desired extension to this value through the
existing analytic continuations of the gamma functions themselves.
Moreover,
every result displays a  single pole at $s=-1$ in the gamma function in the last
factor of each denominator, while the rest of the expression is finite
at this point. Therefore, each expression has a zero of order one at $s=-1$.

Because Eqs.~\req{ins} show that the two integrals in (\ref{2.21}) have the same
value (even before setting $s=-1$), we conclude, in a neighborhood of $s=-1$,
\beq
\mathcal{E}_{C 1}^{0}(s)= 0.
\eeq

Equation \req{epsc1} shows that
the $\mathcal{E}_{C 1}^{1}(s)$ contribution involves the integration of
the $L_{m 1}^{1}(iv)$ function, determined by Eqs.~\req{Ls}, \req{Deltans}.
In consequence,
\beq
\begin{array}{c}
\ds \mathcal{E}_{C 1}^{1}(s)=-\frac{\hbar}{ 2} \, \frac{s a^{s-1} }{ 2\pi^2} \
\B\left( \frac{1 }{ 2}, 1-\frac{s }{ 2} \right)
\sin\left( \pi\frac{s}{ 2} \right) \\
\ds \times \sum_{m=-\infty}^{\infty}\int_{0}^{\infty} dv \, v^{2-s}
\left[ I_m'(v)K_m'(v)-\left( 1+\frac{m^2}{ v^2} \right) I_m(v)K_m(v)
+\frac{1 }{ v}(I_m(v)K_m(v))' \right].
\end{array}
\label{epsc1i}
\eeq
Bearing in mind the useful property
\be
\frac{d }{ dv}[ v^2 I_m'(v) K_m'(v) -(v^2+m^2) I_m(v) K_m(v) ]=
-2v I_m(v) K_m(v),
\ee
we apply partial integration omitting a `boundary term' which
vanishes for a given $s$ range ($2 < \Re s < 3$). Doing so, we find that
Eq.~\req{epsc1i} becomes
\beq
\begin{array}{c}
\ds \mathcal{E}_{C 1}^{1}(s)=-\frac{\hbar}{ 2} \, \frac{s a^{s-1} }{ 2\pi^2} \
\B\left( \frac{1 }{ 2}, 1-\frac{s }{ 2} \right)
\sin\left( \pi \frac{s}{ 2}  \right) \\
\ds\times\left[
\int_{0}^{\infty} dv \, v^{1-s} \sum_{m=-\infty}^{\infty} (I_m(v) K_m(v))'
+\frac{2 }{ 1-s}\int_{0}^{\infty} dv \, v^{2-s} \sum_{m=-\infty}^{\infty} I_m(v) K_m(v)
\right] .
\end{array}
\label{api}
\eeq
Because of this $s$ restriction, the integrals in \req{api} cannot be
directly taken at $s= -1$.
However, if this difficulty is disregarded,
we may formally set $s= -1$ and find
\beq
\mathcal{E}_{C 1}^{1}(s=-1) \to
-\frac{\hbar }{8 \pi a^2} \int_{0}^{\infty} dv \, v^2
\sum_{m=-\infty}^{\infty} (I_m(v) K_m(v))'
-\frac{\hbar }{ 8 \pi a^2}\int_{0}^{\infty} dv \, v^3
\sum_{m=-\infty}^{\infty} I_m(v) K_m(v) .
\label{epsc11r}
\eeq
One could argue that the first part can be dismissed
as a mere contact term (since it may be shown
 from Eq.~\req{thrm} that it is local in $v$).
The second part of \req{epsc11r} gives a contribution which exactly cancels the
bulk contribution found in Ref.~\cite{CPM}, where the same type of
formal expressions was employed. [See formulas (72), (78) there.
In that paper what  was evaluated was the Casimir
radial pressure $P_C$, which is related to $\mathcal{E}_C$
through Eq.~\req{PepsC}.]

Perhaps a better argument is to again use the technique of multiplying each term
in the $m$ summation by $1= -v W[I_m(v),K_m(v)]$, and turn the initial expression
into a linear
combination of integrals with summations of products of four Bessel
functions [like Eq.~\req{ins}]. That linear combination would yield an
identically null result --- one that is zero for any $s$
value --- by virtue of the symmetries seen in Eqs.~\req{ins} under interchange
of different Bessel function types
(see also comment after Eqs.~(80) in Ref.~\cite{CPM}).
Hence, near $s=-1$,
\beq
\mathcal{E}_{C 1}^{1}(s)= 0.
\label{2.26}
\eeq

From another viewpoint,
according to the reasonings in Ref.~\cite{LSN} (and references therein),
which dealt with a similar problem for a dielectric ball,
all linear terms in $(\varepsilon_2-\varepsilon_1)$ have
to be subtracted because they represent the self-energy of the
electromagnetic field due  to polarizable particles. Therefore,
one simply must remove the linear part, no matter what precise form it
has.  This, of course, is the physical basis for removing the bulk energy
contribution.

\subsection{Second order in $(\varepsilon-1)$}

We start with the part called $\mathcal{E}_{C 2}^{20 A}(s)$,
because its evaluation is most similar to that of the $\mathcal{E}_{C
1}^0(s)$, $\mathcal{E}_{C 1}^1(s)$ contributions.
Using the fourth line in Eq.~\req{epsc2},
the fifth in Eq.~\req{Ls}, expressions \req{Deltans} with $y=iv$,
introducing, again, a unit factor in terms of the Wronskian,
and following the same argument that led to (\ref{2.26}), one arrives at
\beq
\mathcal{E}_{C 2}^{20 A}(s)= 0,
\label{epsc20A}
\eeq
near $s=-1$.
Next, selecting the lines in Eqs.~\req{epsc2},
which define
$\mathcal{E}_{C 2}^{00}(s)$,   $\mathcal{E}_{C 2}^{10}(s)$,
$\mathcal{E}_{C 2}^{20 B}(s)$, $\mathcal{E}_{C 2}^{11}(s)$,
the parts of Eq.~\req{Ls} which determine
$L_{m 2}^{00}(y)$, $L_{m 2}^{10}(y)$,
$L_{m 2}^{20 B}(y)$, $L_{m 2}^{11}(y)$,
the form of
$\Delta_m^{(1,0)}(y)$ given by \req{Deltans}
(its square for the case of $L_{m 2}^{20 B}(y)$),
and taking imaginary arguments $y= iv$, one finds
\begin{subequations}
\label{epsc2s}
\bea
\ds \mathcal{E}_{C 2}^{00}(s)&=&\ds
\frac{\hbar }{ 2} \, \frac{ s a^{s-1} }{ 4\pi^2 } \
\B\left( \frac{1 }{ 2} , -\frac{s }{ 2} \right)
\sin\left( -\pi \frac{s }{ 2} \right)
\int_{0}^\infty dv \, v^{2-s} \sum_{m=-\infty}^{\infty}
{I_m'}^2(v) K_m^2(v), \label{epsc2sa}
\eea
\bea
\ds \mathcal{E}_{C 2}^{10}(s)&=&\ds
\frac{\hbar }{ 2} \, \frac{ s a^{s-1} }{ 4\pi^2 } \
\B\left( \frac{1 }{ 2} , 1-\frac{s }{ 2} \right)
\sin\left( \pi \frac{s }{ 2}  \right) \nonumber\\
&&\quad
\ds \times \int_{0}^\infty dv \, v^{2-s} \sum_{m=-\infty}^{\infty} \ds \bigg[
2 I_m(v) I_m'(v) K_m(v) K_m'(v)
+v \, {I_m'}^2(v) K_m(v) K_m'(v) \nonumber \\
&&\qquad\ds -\left( v + \frac{m^2 }{ v} \right) I_m^2(v) K_m(v) K_m'(v)
\bigg] ,
 \label{epsc2sb}
 \eea
 \bea
\ds\mathcal{E}_{C 2}^{20 B}(s)&=&\ds
\frac{\hbar }{ 2} \, \frac{ s a^{s-1} }{ 8\pi^2 } \
\B\left( \frac{1 }{ 2} , 2-\frac{s }{ 2} \right)
\sin\left( -\pi \frac{s }{ 2}  \right) \nonumber\\
&&\quad
\ds
\times \int_{0}^\infty dv \, v^{2-s} \sum_{m=-\infty}^{\infty} \ds \bigg[
{I_m'}^2(v) K_m^2(v) + I_m^2(v) {K_m'}^2(v) \nonumber\\
&&\qquad\mbox{}+2(1-v^2-m^2)I_m(v) I_m'(v) K_m(v)
K_m'(v) \nonumber \\
&&\qquad\ds \mbox{}+v^2 {I_m'}^2(v) {K_m'}^2(v) +\left( v^2+2m^2+\frac{m^4}{ v^2}
\right) I_m^2(v) K_m^2(v)  \nonumber\\
&&\qquad\ds  \mbox{}+2v \, {I_m'}^2(v) K_m(v) K_m'(v) + 2v \, I_m(v) I_m'(v)
{K_m'}^2(v) \nonumber\\
&&\qquad\mbox{}-2\left( v + \frac{m^2 }{ v} \right) I_m(v) I_m'(v) K_m^2(v)
-2\left( v + \frac{m^2 }{ v} \right) I_m^2(v) K_m(v) K_m'(v)
\bigg] ,\nonumber\\ \label{epsc2sc}
\eea
\bea
\ds\mathcal{E}_{C 2}^{11}(s)&=&\ds\frac{\hbar }{ 2} \, \frac{ s a^{s-1} }{
2\pi^2 } \ \B\left( \frac{3 }{ 2} , 1-\frac{s }{ 2} \right)
\sin\left( \pi \frac{s }{ 2}  \right)
\int_{0}^\infty dv \, v^{-s} \sum_{m=-\infty}^{\infty} m^2 I_m^2(v) K_m^2(v) .
\nonumber\\
\label{epsc2sd}
\eea
\end{subequations}
The $v$ integrals are again evaluated in the manner illustrated in
 Eq.~\req{ins}. All of the resulting formulas  exhibit
zeros of order one at $s=-1,-3,-5,\dots$.
Since all the beta and sine functions in Eq.~\req{epsc2s} are finite at
$s= -1$, each of these quantities will just yield
${\cal O}(s+1)$. In fact, after calculating the coefficients,
\begin{subequations}
\label{epsc2sr}
\bea
\mathcal{E}_{C 2}^{00}(s)&=&\ds
\hbar \frac{1 }{ 192 \, \pi \, a^2}(s+1)
+{\cal O}((s+1)^2),\label{epsc2sra} \\
\mathcal{E}_{C 2}^{10}(s)&=&\ds
-\hbar \frac{1 }{ 288 \, \pi \, a^2}(s+1)
+{\cal O}((s+1)^2), \label{epsc2srb}\\
\mathcal{E}_{C 2}^{20 B}(s)&=&\ds
\hbar \frac{7 }{ 3840 \, \pi \, a^2}(s+1)
+{\cal O}((s+1)^2), \label{epsc2src}\\
\mathcal{E}_{C 2}^{11}(s)&=&\ds\hbar\frac{1 }{2304 \, \pi \, a^2}(s+1)
+{\cal O}((s+1)^2).\label{epsc2srd}
\eea
\end{subequations}
From Eqs.~\req{epsc20A} and \req{epsc2sr}, we see that
$\ds\lim_{s \to -1}\mathcal{E}_{C 2}(s)=0$, i.e.,
the $(\varepsilon-1)^2$ contribution to the Casimir energy
per unit length in the dilute-dielectric approximation is zero.

\section{Concluding remarks}
Applying an analytic regularization which changes the
eigenmode power, we have calculated --- to the order of
$(\varepsilon-1)^2$ ---
what in Ref.~\cite{barton} is called a pure Casimir term, i.e., the
convergent component of the energy depending only on $\hbar, c$,
the electrostatic polarizability of the material and the
dimensions of the body.
Using the  words of Ref.~\cite{F},
a forest of gamma functions has grown out of
an analytic continuation. Although the applied technique might be
regarded as somewhat physically opaque, its relations to more transparent
regularizations have already been studied (see e.g. Refs.~\cite{CVZ}).
In Ref.~\cite{CPM} the only relevant contribution of the
bulk part took place at the order of $(\varepsilon-1)^1$.
Since in the present regularization the corresponding term
vanishes by itself, to omit or include a separate bulk part
leaves the outcome unchanged. This remark is true as far as the
analytic regularization method proposed in Sec.~6.1 of Ref.~\cite{CPM} is
concerned, but not so for the (regulated) numerical method presented in
Secs.~6.2 and 6.3 there, where a detailed numerical cancellation occurs
between the bulk and cylinder parts.
As a result, the pure Casimir term is seen to vanish
to order $(\varepsilon-1)^2$.
Comparing this calculation to the one offered in Ref.~\cite{CPM},
a decrease in difficulty can be appreciated in the derivation of the
expression to be evaluated.
However, we should stress that a substantial part of this merit is not in
the nature of the method itself, but in the use of a previously known
equation for the classical modes of this problem.
Another key ingredient has been the exploitation of a
Bessel function addition theorem \cite{KR,LSN,CPM}.

Ref.~\cite{BP} exhibits the presence of a divergence
at the order of $(\varepsilon-1)^3$.
Among all the possible divergences which show
up by point-splitting or ultraviolet cutoffs, the ones
which survive in analytic regularization schemes
introduce an unavoidable ambiguity
parametrized by an arbitrary length or mass scale.
This issue is
viewed with more or less concern, depending on the authors
(see comments in Ref.~\cite{F}). The standpoint of Ref.~\cite{BP} was to
admit that the problem is simply ill-defined because its posing
constitutes an idealization (the permittivity $\varepsilon$
as a function of the radial coordinate is treated as a step function).

In any case, the reason why the Casimir energy term of order $(\varepsilon
-1)^2$ is zero for the cylinder, while it is finite for other geometries,
remains rather mysterious. It is clear that we still have some way to go
to understand quantum vacuum energies.
\appendix

\acknowledgments{

 A.R. wishes to thank V.V. Nesterenko for kind correspondence
at the early stages of this work.  K.A.M. acknowledges useful conversations
with In\'es Cavero-Pel\'aez.  The work of K.A.M. is supported in part by
a grant from the U.S. Department of Energy.}

\end{document}